# Insights into the Kinematic Evolution of Filament-induced Jets: Coupling of Interchange and Flare Reconnection

Zehao Tang (汤泽浩)[1,2], Xiaoli Yan (闫晓理)[1,2], Zhike Xue (薛志科)[1,2], Jincheng Wang (王金成)[1,2], Zhe Xu (徐喆)[1,2], and Yadan Duan (段雅丹)[1,2]
[1] Yunnan Observatories, Chinese Academy of Sciences, Kunming, 650216, People's Republic of China; yanxl@ynao.ac.cn
[2] Yunnan Key Laboratory of Solar Physics and Space Science, Kunming 650216, People's Republic of China
*Received 2026 July 30; revised 2026 September 16; accepted 2026 September 16; published 2026 October 28*

## Abstract

Jets are ubiquitous eruptive activities in the solar atmosphere, playing an important role in mass transport, atmospheric heating, and energy supply. However, the triggering mechanism remains elusive. To reveal the fine process of jet evolution, we study a jet associated with a B-class flare that involves in two distinct stages. Analyzing the complete evolution of a solar jet captured by the Solar Dynamics Observatory (SDO), we clearly identify its dual-stage kinematics, evolving from interchange to flare reconnection. This dual-stage jet is characterized by distinct dual-stage signatures in magnetic morphology, H$\alpha$ profile, jet speed, jet height, X-ray flux, loop-top source, and temperature. These differences are naturally unified by a physical picture coupling interchange and flare reconnection, offering new insights into understanding of the physical process of filament-induced jets.

## 1. Introduction

Jet activities, manifesting as collimated plasma beams across an extremely wide range of spatial scales (from subarcsec to hundreds of arcsec), pervade the solar atmosphere (K. Shibata et al. 1994, 2007; M. Shimojo & K. Shibata 2000; A. R. Paraschiv et al. 2015; J. Liu et al. 2023). They are ubiquitous, occurring in all types of solar regions—including active regions, coronal holes, and quiet-Sun regions—and continuously supply mass and energy to the upper atmosphere. Consequently, they represent one of the most important sources for heating coronal plasma and accelerating the solar wind (e.g., G. E. Brueckner & J.-D. F. Bartoe 1983; J. W. Cirtain et al. 2007; H. Tian et al. 2014; L. P. Chitta et al. 2023) and provide deep insight into solar reconnection (e.g., Q. M. Zhang & H. S. Ji 2014; L. Ni et al. 2021; G. Cheng et al. 2024; D. Nóbrega-Siverio et al. 2025; Y. Wang et al. 2025; Y. Duan et al. 2026a; Y. Liu et al. 2026; Z. Tang 2026).

Based on morphology, R. L. Moore et al. (2010) proposed a dichotomy of solar coronal jets: standard jets and blowout jets. Standard jets typically appear as simple, narrow, well-collimated inverted-Y structures without significant twist release, whereas blowout jets exhibit broader and more complex morphologies associated with untwisting motions (R. L. Moore et al. 2010, 2013; C. Liu et al. 2011; A. C. Sterling et al. 2022; Y. Chen et al. 2026). Beyond morphological differences, the two jet types differ significantly in several physical aspects. Compared with the standard jet stage, the blowout jet stage exhibits higher plasma temperatures (J. Hong et al. 2017), higher reconnection rates (C. Liu et al. 2011), more pronounced X-ray flux enhancements (C. Liu et al. 2011; R. Chandra et al. 2017), and additionally, a high-speed component. This high-speed component corresponds to Alfvén waves originating from the untwisting of magnetic field lines (J. W. Cirtain et al. 2007; E. Pariat et al. 2016; D. M. Long et al. 2023; Y. Chen et al. 2026). Collectively, these differences indicate that blowout jets involve more intense energy release than standard jets. The significant differences between standard and blowout jets reflect their distinct formation mechanisms. A widely accepted scenario is that standard jets and blowout jets arise from interchange reconnection between open and closed loops and from flare reconnection triggered by erupting filaments or flux ropes, respectively.

An increasing number of solar jet studies have reported their dual-stage evolution processes. (C. Liu et al. 2011; V. Archontis & A. W. Hood 2013; E. Pariat et al. 2016; X. Zhu et al. 2017; S. Mandal et al. 2022; Y. Chen et al. 2026). X. Zhu et al. (2017) reported a blowout jet that develops closely from a standard one. C. Liu et al. (2011) conducted an observational study of a standard-to-blowout jet, showing that the standard and blowout jet stages are driven by interchange reconnection and subsequent filament eruption, respectively. The erupting filament is identified as a miniature version of a coronal mass ejection (CME), as its eruption shares a similar mechanism with large-scale CMEs (R. L. Moore et al. 2010; C. Liu et al. 2011; A. C. Sterling et al. 2015; P. F. Wyper et al. 2017). V. Archontis & A. W. Hood (2013) further reported that open–close reconnection is coupled with internal arcade reconnection in standard-to-blowout jets. These studies suggest distinct kinematic evolution in solar jets. The details from the numerical results of F. Fang et al. (2014) and Y. Chen et al. (2026) reveal successive occurrence of interchange and flare reconnection in solar jets, offering insights into their distinct evolution stages. However, the physical link bridging the standard and blowout stages of jets remains debated, and existing theories yield conflicting results. C. Liu et al. (2011) and R. Chandra et al. (2017) suggested that magnetic cancellation may drive the transition from the standard to the blowout stage. On the contrary, magnetic emergence is widely evidenced as a common driver of successive standard and blowout jets in numerical studies (T. Török et al. 2009; V. Archontis & A. W. Hood 2013; F. Fang et al. 2014; Y. Chen et al. 2026). Moreover, jet types show a dependence on the inclination of magnetic fields (E. Pariat et al. 2015).

Although many observations have reported dual-stage evolution in solar jets, evolution details are often difficult to resolve due to complexity added by complex magnetic fields, higher plasma temperatures, and more dynamic multithermal structures. Here, we present a detailed study of a filament-induced jet, which provides clear morphological evidence for the successive occurrence of interchange and flare reconnection, matching with numerical results. Detailed analysis focusing on the physical link between interchange and flare reconnection emphasizes the fundamental role of interchange reconnection in triggering flare reconnection, shedding light on their coupling process. Section 2 describes observations and methods, Section 3 presents our detailed results, and Section 4 provides a comprehensive conclusion and discussion.

## 2. Observations and Methods

This event originated from active region NOAA 12144 on 2014 August 17. In the duration, the GOES 15 1–8 Å flux peaked twice, at approximately 06:52 UT and 07:03 UT, respectively. The later peak was registered by GOES 15 as a B8 class flare in the 1–8 Å band and by the Reuven Ramaty High Energy Solar Spectroscopic Imager (RHESSI; R. P. Lin et al. 2002) in the 3–6 and 6–12 keV energy ranges. We used the Atmospheric Imaging Assembly (AIA; J. R. Lemen et al. 2012) and the Helioseismic and Magnetic Imager (HMI; J. Schou et al. 2012) on board the Solar Dynamics Observatory (SDO) to trace the evolution of the photospheric magnetic field and the coronal magnetic topology. SDO/HMI images photospheric line-of-sight (LOS) magnetic fields with a 45 s cadence and 0.″5 pixel resolution. SDO/AIA EUV channels observe coronal activities with a 12 s cadence and a 0.″6 pixel resolution. We also used H$\alpha$ (6563 Å) observations from the Global Oscillation Network Group (GONG), which provides H$\alpha$ full-disk observations with a pixel scale and cadence of 1″ and ∼1 minute (J. W. Harvey et al. 2011).

The fifferential emission measure (DEM) derived from six AIA EUV channels (94, 131, 171, 193, 211, and 335 Å) was used to diagnose plasma temperatures and number densities. The DEM is determined by the following equation:

$$I_i = \int R_i(T) \times \mathrm{DEM}(T) dT, \tag{1}$$

where $I_i$ is the observed emission intensity of a wavelength channel $i$; $R_i(T)$ is the temperature response function of the channel $i$; DEM($T$) represents the DEM of coronal plasma, computed through the method of I. G. Hannah & E. P. Kontar (2012). After solving the DEM, the emission measure (EM) within a temperature range ($T_{\min}$, $T_{\max}$) is calculated through the following equation:

$$\mathrm{EM} = \int_{T_{\min}}^{T_{\max}} \mathrm{DEM}(T) dT. \tag{2}$$

Finally, the weighted temperature is obtained through the EM and DEM

$$\bar{T} = \frac{\int_{T_{\min}}^{T_{\max}} \mathrm{DEM}(T) \times T dT}{\mathrm{EM}}, \tag{3}$$

where $\bar{T}$ is the weighted temperature (X. Cheng et al. 2012). The average electron number density $n_e$ is estimated according to $\sqrt{\mathrm{EM}/H}$ (assuming filling factor ≈ 1), where $H$ is the depth along the LOS. In this study, we assumed the LOS depth as ∼3 Mm.

## 3. Results

### 3.1. Dual-stage Jet

We focus on a filament-induced jet that occurred on 2014 August 17, in active region NOAA 12144, with heliocentric coordinates of approximately (740″, −360″). All observational images have been rotated by 90° so that the jet is directed upward. As time progresses, the event is characterized by the following two distinct stages.

*Interchange reconnection stage*. Figure 1 displays the first stage of this event as observed by SDO/AIA 304 Å (left column), SDO/AIA 131 Å (middle column), and SDO/HMI LOS magnetogram superimposed by 131 Å images (right column), with a field of view of 70″ × 70″. Contours are overlaid on the SDO/HMI images to highlight the motion of magnetic polarities. The time sequence runs from top to bottom. Initially (∼06:00 UT), a filament lies in the source region, appearing as dark threads (see Figures 1(a) and (b)) rooted in a mixed-polarity region (labeled P1 and N1; see Figure 1(c)). Due to resolution limitations, the magnetic connectivity is difficult to discern. Initially, no prominent atmospheric features are visible in the source region, apart from a few compact brightenings with faint EUV emission (see Figures 1(a) and (b)).

The filament connectivity at about 06:47 UT is more distinguishable than before. Figure 1(d) clearly reveals that one side of the filament was opened, and upward-ejected filament material forms a jet. This filament and jet are hardly visible in the SDO/AIA 131 Å channel (Figure 1(e)); however, this channel reveals two representative magnetic topologies: an overlying Y-point and an underlying loop (Figure 1(e)). These are typical products of interchange reconnection (i.e., open–close reconnection) occurring at the interface between a closed loop and an open ambient field of opposite polarity, widely seen in solar jets (K. Shibata et al. 2007; H. Tian et al. 2014, 2018). According to the interchange reconnection picture, we identify that the two structures correspond to a magnetic cusp and a post-reconnection loop, respectively. At the later stage, the jet connects to a remote brightening that is unambiguously rooted in a negative-polarity quiet-Sun field. Based on this connectivity, before the footpoint exchange by the interchange reconnection, the ambient field involved in the reconnection has its near-side footpoint rooted in P1. Therefore, the initially closed filament is rooted in N1-P1. We depict this pre-reconnection configuration in Figure 1(f). The interchange reconnection occurs at the interface between N1 and P1 (Figure 1(f)). Following this reconnection, the footpoint of the filament is exchanged from P1-N1 to P1-remote negative polarity, i.e., the remote footpoint of the long ambient loop, resulting in the

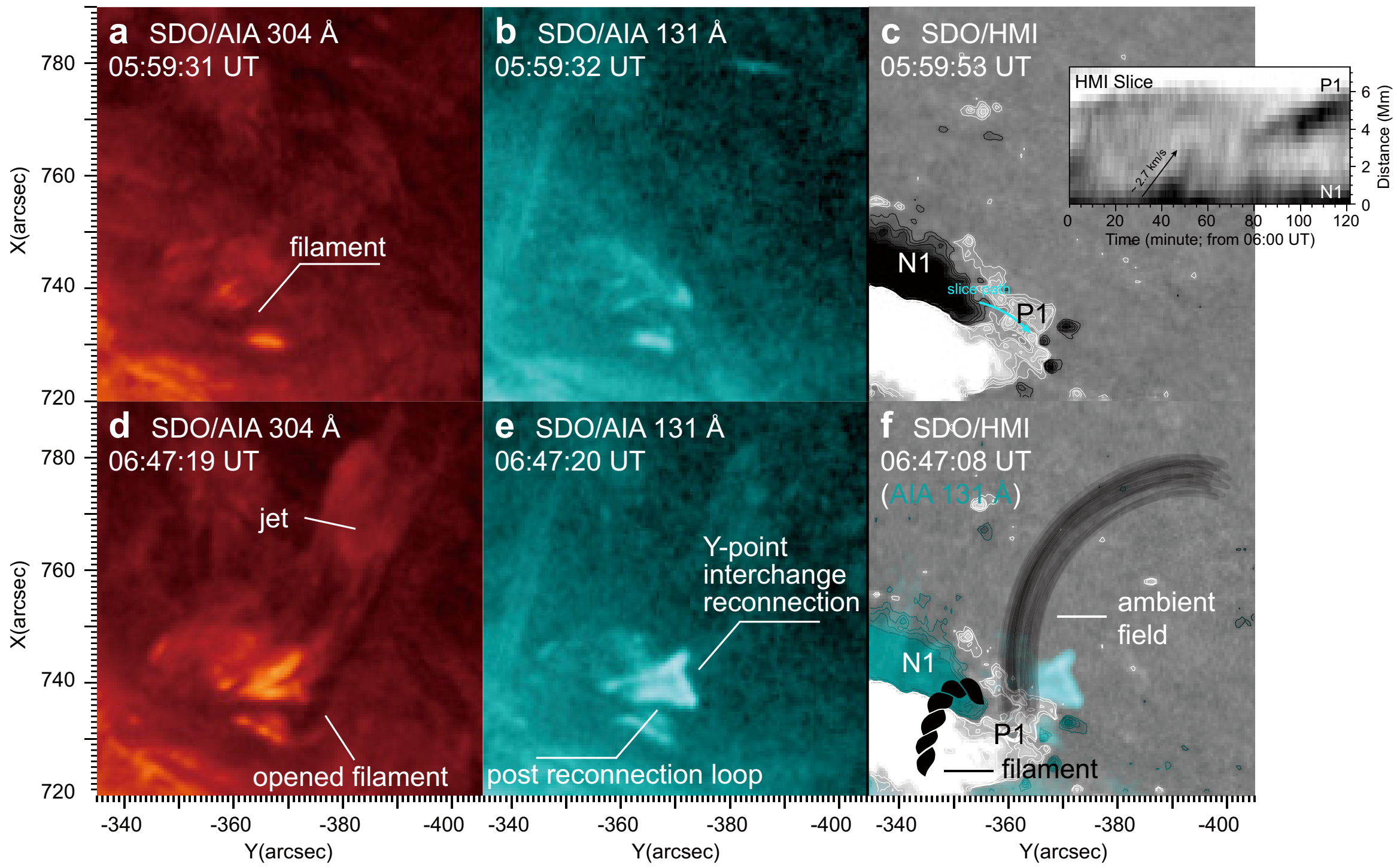

**Figure 1.** Multiwavelength diagnosis of the interchange reconnection stage. Left to right columns: SDO/AIA 304 Å SDO/AIA 131 Å and SDO/HMI observations. Time runs top to bottom. Contours are overlaid on SDO/HMI to guide eyes. The top-right HMI slice in panel (c) shows temporal evolution of P1 and N1 along the cyan curve between the two polarities. Cartoon versions of the filament and ambient fields are depicted in panel (f) to show pre-reconnection magnetic configuration. Representative structures and information are labeled in the corresponding panels, including filament (panel (a)), key magnetic polarities (panels (c) and (f)), jet and opened filament (panel (d)), and Y-point/interchange reconnection and post-reconnection loop (panel (e)). The animation covers the time interval from 06:00 to 07:00 UT, with a real-time duration of 10 s and a temporal cadence of 12 s. It shows the continuous evolution of SDO/AIA 304 Å, 131 Å, SDO/HMI line-of-sight magnetograms, DEM electron-density maps, DEM weighted-temperature maps, and GONG Hα images. These data depict the evolutionary sequence starting from the initial configuration, through the interchange reconnection jet, to the subsequent filament eruption and flare reconnection jet.

(An animation of this figure is available in the online article.)

opening filament and the jet seen in Figure 1(d), as well as the magnetic cusp and post-reconnection loops seen in Figure 1(e). Given that the ambient field is a large-scale, distantly closing coronal loop, the term "interchange reconnection" used here is not strictly open–closed reconnection; it denotes a connectivity exchange between the short closed filament and a much longer coronal loop. The slight spatial offset between the observed reconnection productions and the schematic in Figure 1(f) is likely due to projection effects, because the event is close to the western limb.

The time slice of SDO/HMI magnetograms in the top-right corner shows the evolution of P1 and N1. We identify the motion of N1 toward P1 at an average speed of about 2.7 km $s^{-1}$ during the interchange reconnection stage. Such convergence of opposite polarities, which brings opposite magnetic field lines together, plays a fundamental role in facilitating the formation of the reconnecting current sheet (C. Liu et al. 2011; N. C. Joshi et al. 2018; P. Kumar et al. 2018; E. R. Priest et al. 2018; R. Joshi et al. 2020; J. Yang et al. 2024; F. Moreno-Insertis et al. 2025; Z. Tang et al. 2025a, 2025b), and is therefore a potential trigger for the observed interchange reconnection.

*Flare reconnection stage*. After the interchange reconnection, the evolution of this event enters the flare reconnection stage at about 06:57 UT, as shown in Figure 2, whose field of view (FOV) and layout are the same as in Figure 1. At this stage, the opened side of the filament rises at about 06:57 UT (see Figures 2(a) and (b)), while the unopened side remains anchored on P1 (Figure 2(c)). As a result of its continued ascent, this filament erupts (∼07:02 UT) and generates a jet that is wider and brighter than before. The emission of the jet base is intensified, suggesting more violent magnetic energy release. Behind the filament, an X-point appears, and we outlined the configuration of the erupting filament in Figure 2(g) to show its magnetic connections. This X-point connects the upward erupting filament as well as an underneath loop that is rooted in N1-P1. Such a magnetic configuration is analogous to that of the flare reconnection picture (S. Masuda et al. 1994; M. Ohyama & K. Shibata 1998; J. Lin & T. G. Forbes 2000; W. Liu et al. 2013; X. L. Yan et al. 2018; Y. Chen et al. 2022), in which legs of Ω-shaped confined fields of the filament converge as a result of the filament's ascent, giving rise to a reconnecting X-point at the separator between legs. The reconnecting upflow flux further facilitates the filament erupting, with downflow flux forming post-reconnecting loops connecting the photosphere. Closely following the appearance of this X-point, a loop-top source in the RHESSI 6–12 keV range is registered at exactly the same spatial position, as suggested by the comparison of Figures 2(g) and (h). Based on these observational indicators, we identify that flare reconnection occurs at the X-point. This flare reconnection drives the filament eruption, the formation of the jet, the postflare loop, and the loop-top source.

Generally, the flare reconnection in solar jets can be attributed to the sustained rise of the filament. However, the cause of this sustained filament rise is not unique. On the one hand, photospheric magnetic shearing and flux emergence often lead to a continuously increasing magnetic pressure, which has been reported as a facilitator of the sustained filament rise and serves as a common trigger (J. T. Karpen et al. 2012; V. Archontis & A. W. Hood 2013; P. F. Wyper et al. 2017; C. Jiang et al. 2021). On the other hand, the sustained filament rise has also been closely associated with interchange reconnection, likely due to sustained removal of confined field overlying filaments (R. L. Moore et al. 2010; Y. Chen et al. 2026). In the animated Figure 1, no significant shearing, emergence, or cancellation is evident in the photospheric magnetic flux during the flare reconnection stage. Considering this, we suggest that the filament rise in this event is more likely related to the interchange reconnection. Further details will be presented in Section 3.4.

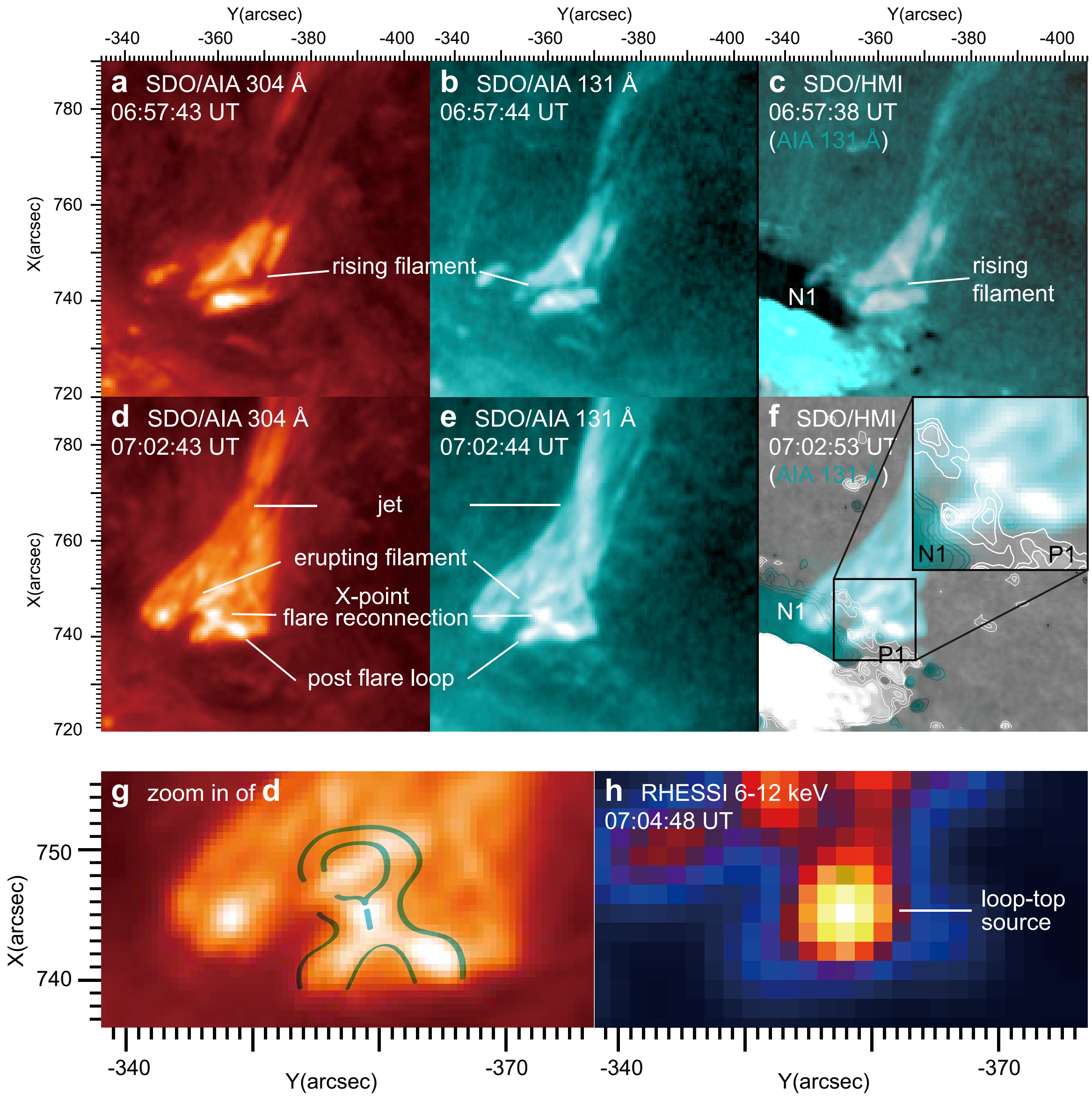

**Figure 2.** Multiwavelength diagnosis of the flare reconnection stage. The layout and FOV are kept the same as Figure 1. Panels (g) and (h) are zoom-ins of panel (d) and its counterpart in RHESSI 6–12 keV; an outline is added to panel (g) to highlight the reconnection configuration. Representative structures and information are labeled in the corresponding panels, including key magnetic polarities (panels (c) and (f)), the rising filament (panels (a), (b) and (c)), the jet, the erupting filament, the X-point/flare reconnection and post flare loop (panels (d) and (e)), and the loop-top source (panel (h)).

### 3.2. Hα Observations and DEM Diagnosis

Figure 3 presents the Hα and DEM diagnosis of the two jet stages. At the interchange reconnection stage, the jet appears as upward-ejected dark, cool material in the Hα observations, a phenomenon alternatively referred to as a solar surge. Distinct from the SDO/AIA observations shown in Figure 1, the jet base remains completely invisible in the Hα channel.

Electron number density and weighted temperature maps derived from the DEM method is used to further investigate the fine structure of the jet base at this stage. Clearer fine structures are discernible in the density map, where the jet base consists of two Y-points (Figure 3(b)). These two Y-points, also known as magnetic cusps, represent the reconnection outflow regions. The plasma sheet connecting the two Y-points is the reconnecting current sheet (Figure 3(b)), which is responsible for the interchange reconnection identified in Figure 1.

The temperature profile revealed in Figure 3(c) suggests that heated areas are spatially confined. A comparison of the temperature and density maps reveals the correspondence of high-temperature regions ($T \geqslant 10^{6.7}$ K) with two Y-points.

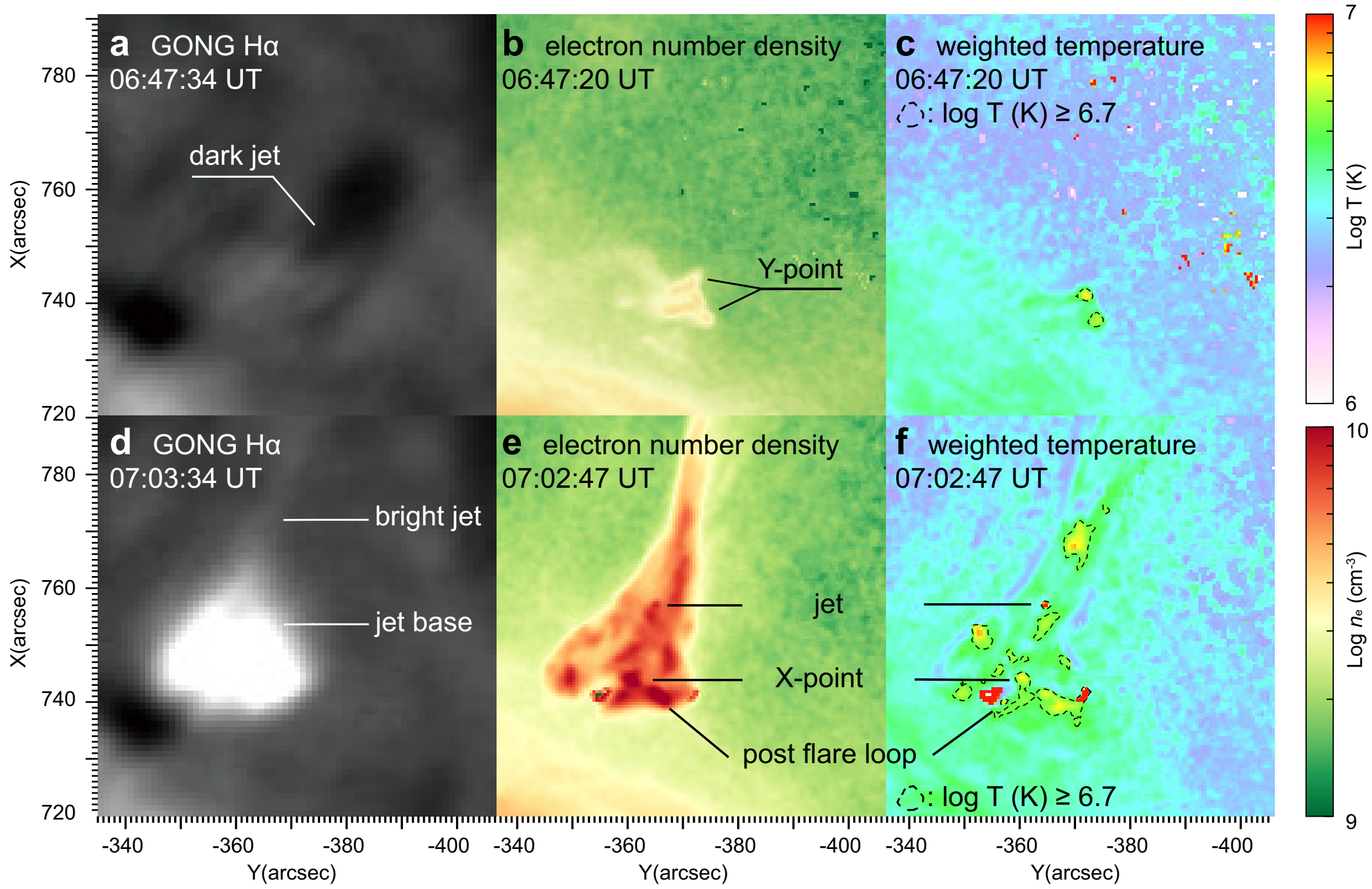

**Figure 3.** Hα and DEM diagnosis of the dual-stage jet. The FOV is kept the same as Figure 1. Left to right columns: GONG Hα, DEM electron number density ($n_e$), and DEM weighted temperature ($T$) maps. Time runs top to bottom. Representative structures and information are labeled in the corresponding panels, including the dark jet (panel (a)), the Y-point (panel (b)), the bright jet and jet base (panel (d)), and the jet, X-point, and postflare loop (panels (e) and (f)). Areas with $T \geqslant 10^{6.7}$ K are contoured in panels (c) and (f).

Similar observational results have also been reported by Z. Tang (2026), L. Li et al. (2016), L. Yang et al. (2026), and J. Hong et al. (2025), which clearly show that the cusps are hotter than the current sheets. The underlying reason is that Y-points serve as primary sites for energy deposition, where multiple heating mechanisms, including shocks, betatron processes, and Fermi acceleration, can operate (T. G. Forbes & E. R. Priest 1983; S. Masuda et al. 1994; F. Guo & J. Giacalone 2012; W. Liu et al. 2013; M. Liu et al. 2023; B. Chen et al. 2024). A close inspection reveals a notable difference: the upper Y-point is hotter than the lower Y-point. This asymmetry can be naturally attributed to the heating of the lower Y-point occurring in a lower layer with denser gas.

The flare reconnection stage exhibits the following features that are significantly distinct from the preceding stage. In the Hα image, a bright jet is observed, emanating from a luminous jet base, overall in an inverted-Y shape, a typical structure that is often observed in multiscale reconnecting jets (N. Nishizuka et al. 2008; Z. Zeng et al. 2016; R. Chandra et al. 2017; H. Tian et al. 2018). In the Hα image, details are lost due to the compact brightening of the jet base, but the density map delineates finer structures, including a wider jet, an X-point, and a postflare loop (see Figure 3(e)), which are highly similar to those depicted in Figures 2(d) and (e). The temperature map further indicates that high-temperature regions ($T \geqslant 10^{6.7}$ K) are not confined to the flare reconnection region (i.e., the X-point), but are instead distributed throughout the entire jet, encompassing both the jet base and its spire. This suggests a more vigorous energy release process during the flare reconnection stage.

Three key transitions emerge from a comparison of Figures 3(a)–(c) with Figures 3(d)–(f). First, the jet base becomes visible in Hα, accompanied by a transition of the jet itself from dark to bright. Second, the jet becomes detectable in both the density and temperature maps at the flare reconnection stage (Figures 3(e) and (f)), features that are absent in the preceding interchange reconnection stage (Figures 3(b) and (c)). Third, heated areas with temperatures ($T \geqslant 10^{6.7}$ K), delineated by contours, imply a larger-scale heating process. Collectively, these transitions point to a more intense energy release of the flare reconnection stage than the preceding interchange reconnection stage, a finding that will be further explored in a later section.

### 3.3. Dual-stage Energy Release

The previous sections qualitatively identify two stages of the jet that originate from two distinct reconnection stages based on morphological indicators. To quantitatively analyze the energy release process, we construct a temporal–spatial slice using SDO/AIA 304 Å images (Figure 4(a)), together with the simultaneous GOES flux curve (Figure 4(b)) and the temperature curves (Figure 4(c)), to derive key quantities: jet speed, maximum jet height, GOES flux enhancement, and plasma temperature.

In the SDO/AIA 304 Å slice, the temporal–spatial trajectory of the jet appears as two distinct stages in terms of jet speed and jet height (Figure 4(a)). During the interchange reconnection stage, the jet follows a parabolic trajectory, with an average jet speed of $\sim$33 km s$^{-1}$ and a maximum height (relative to $y = 20$ Mm) of $\sim$40 Mm (Figure 4(a)); from 06:34 UT to 06:52 UT). By contrast, during the flare reconnection stage (from 06:52 UT to 07:10 UT), the jet exhibits two higher propagation speeds: 117 and 450 km s$^{-1}$. Moreover, its maximum height exceeds the upper limit of the slice, much greater than the former value of 40 Mm. These distinct kinematic quantities are consistent with the interchange and flare reconnection stages, respectively (Figure 4(a)), indicating that the two reconnection stages are linked to two distinct dynamic processes of the jet.

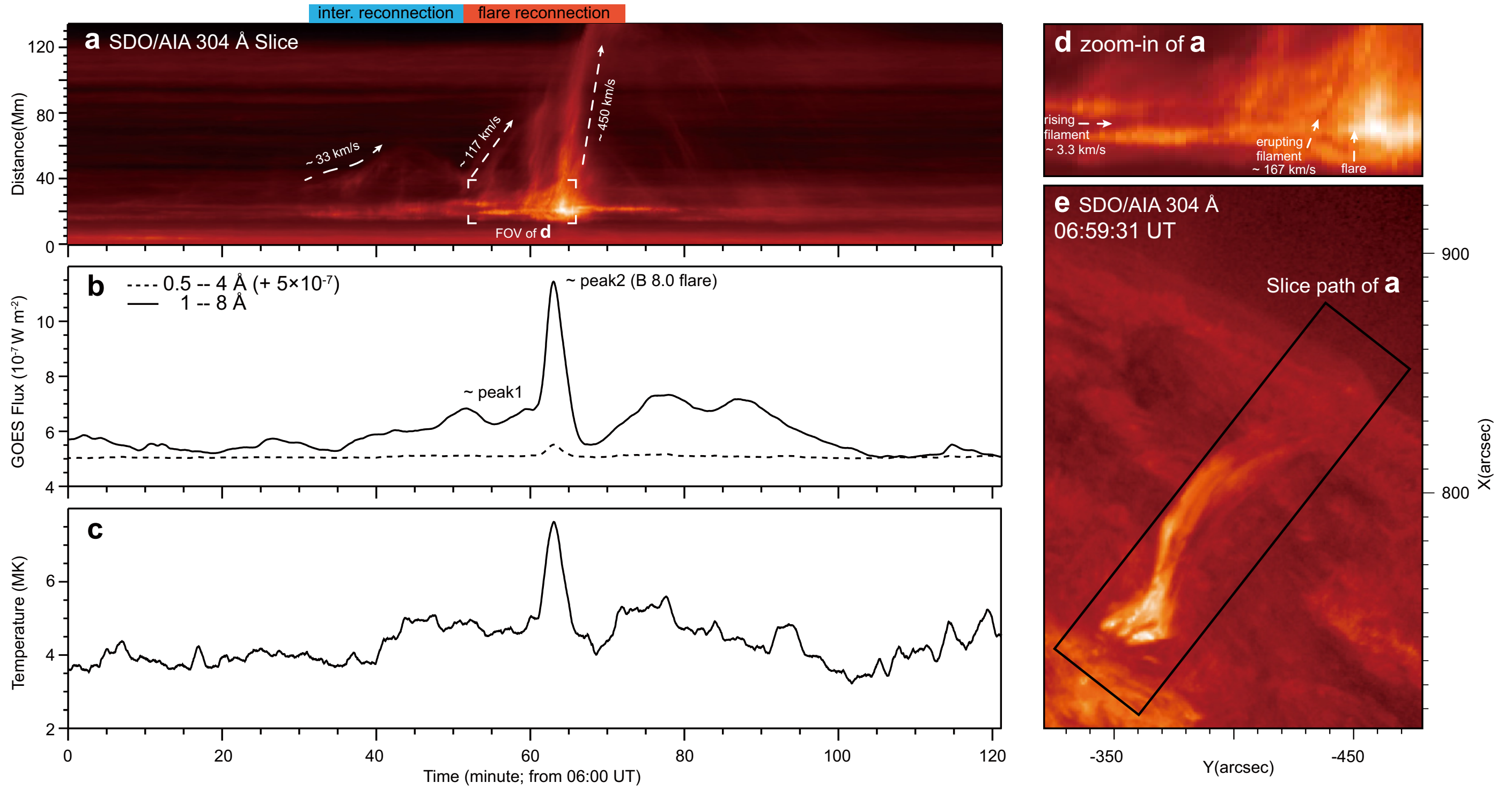

**Figure 4.** Quantitative analysis of dual-stage energy release. Panels (a)–(c) show the SDO/AIA 304 Å slice, simultaneous GOES fluxes, and temperature curves, respectively, with the slice path shown in panel (e). The slice is obtained by averaging the intensity across the width of the rectangular slit in panel (e). Panel (d), a zoom-in of panel (a), reveals the rising and erupting of the filament at the flare reconnection stage.

In agreement with the AIA 304 Å slice, the GOES 1–8 Å flux measurements also show two distinct stages in their enhancement profiles. The first stage, denoted as Peak1 in Figure 4(b), appears as a subtle rise between 06:34 UT and 06:52 UT. The second stage, denoted as Peak2, exhibits a pronounced increase up to a B8.0 flare. These two peaks are respectively consistent with the two reconnection stages.

To further investigate the energy release process of the observed filament-induced jet, we derived the temperature curve based on the GOES 0.5–4 and 1–8 Å flux curves in Figure 4(b) (see S. M. White et al. 2005 for details). The resulting temperature curve is shown in Figure 4(c). The temperature first reaches a plateau of approximately 5 Mk for about 15 minutes (from 06:40 UT to 06:55 UT), and then peaks at 07:03 UT at about 7.5 Mk. These temperature variations are also consistent with the interchange and flare reconnection stages, respectively.

The slice, flux, and temperature analyses, all consistent with two reconnection stages of the observed jet, confirm its property of dual-stage energy release. Specifically, the flare reconnection stage shows significantly higher values in jet speed, maximum height, GOES flux enhancement, and plasma temperature than the interchange reconnection stage. Moreover, Figure 4(d), a zoom-in of Figure 4(a) focusing the flare reconnection stage, reveals a clear evolution of the filament from a slow-rising phase ($\sim$3.3 km s$^{-1}$) to an erupting phase ($\sim$167 km s$^{-1}$), closely followed by the flare. This observation is consistent with the flare reconnection picture shown in Figures 2(d) and (g), suggesting fundamental role of the flare reconnection. In contrast, the interchange reconnection stage features a dark jet (surge; see Figures 1 and 3(a)) and exhibits no eruptive filament. These indicators collectively suggest that flare reconnection involves a more intense energy release than the preceding interchange reconnection, giving rise to distinct jet stages and energy release stages. Similar results can also be seen in C. Liu et al. (2011), Y. Chen et al. (2026), and R. Chandra et al. (2017). The fundamental cause underlying all these differences—including the dual-stage jets, dual-stage reconnection, and dual-stage energy release—is the erupting filament, the details of which are presented in Section 4.

### 3.4. Coupling between Interchange and Flare Reconnection

The sequential cartoons in Figure 5 depict the evolution process of the observed filament-induced jet. Figure 5(a) represents the pre-reconnection configuration, in which an envelope field and a filament are rooted in a bipolar polarity region, with an opposite ambient field rooted adjacently. It should be noted that only the filament is directly observed, whereas the envelope field (alternatively referred to as the confined field) and the ambient field are inferred based on the observational indicators revealed in Figures 1(d) and (e). In this depiction, the filament and the ambient field serve as the closed field and the open field, respectively. At the interface that is sandwiched by the filament and the ambient field often lies the magnetic separatrix, which is a site conducive to magnetic reconnection (E. R. Priest et al. 2018; P. Kumar et al. 2024).

As a result of the footpoint convergence suggested by the HMI slice in Figure 1(c), reconnection occurs at the interface, which produces an overlying Y-point in the reconnection upflow region and a underlying post-reconnection loop in the reconnection downflow region (see Figure 5(b)), corresponding two Y-points seen in Figure 3(b), respectively. Concurrently, the filament opens, with its upward escaping material forming a jet (see Figure 5(b)). These features are all witnessed in Figures 1(d) and (e). The opening filament essentially represents the footpoint interchange of the filament through open–close reconnection (see Figures 1(d), (e), and 3(b)), matching with the interchange reconnection picture (J. Yang et al. 2015; P. F. Wyper et al. 2017, 2018; S. D. Bale et al. 2023; Y. Chen et al. 2026; Y. Duan et al. 2026b). Therefore, the reconnection at this stage is identified as interchange reconnection.

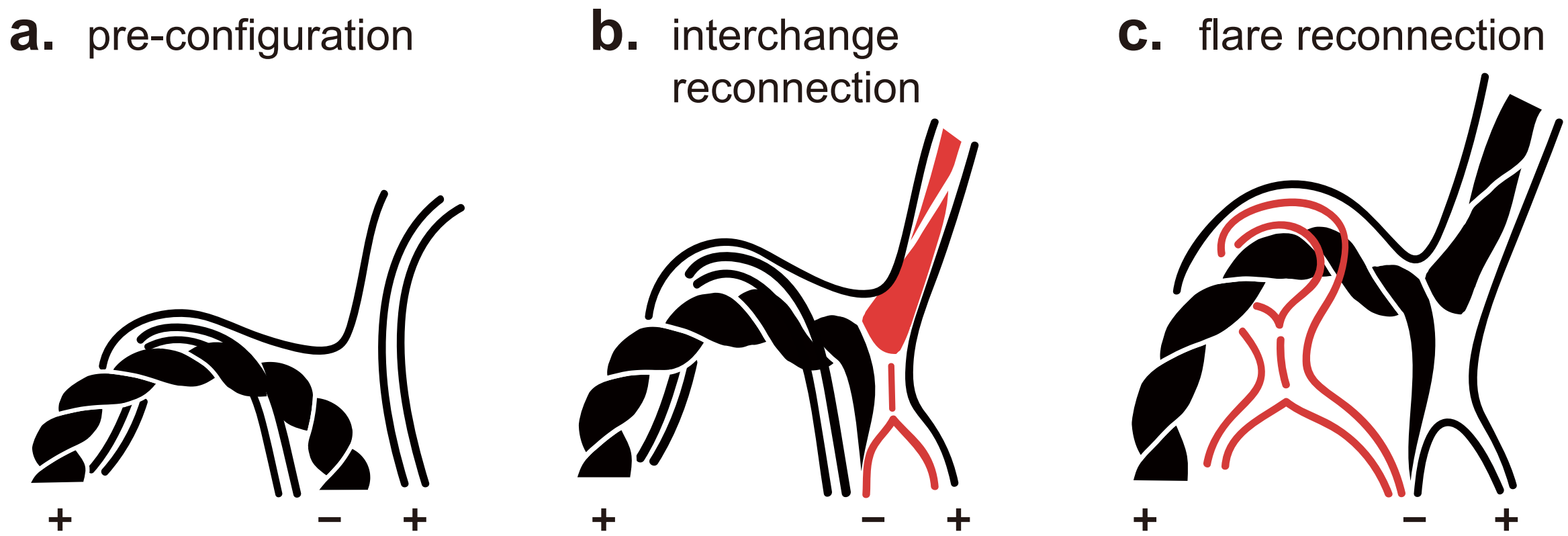

**Figure 5.** Schematic model of dual-stage reconnection in filament-induced jets. Panels (a)–(c) illustrate the evolution from the pre-reconnection configuration (a) to the jet driven by interchange reconnection (b), and finally to the jet driven by flare reconnection (c). In this schematic, the black rope and curves represent the filament and magnetic field lines, respectively. Red lines indicate reconnecting field lines.

Following the interchange reconnection stage, the filament rises, and reconnection occurs at the legs of the envelope field. At the onset of this reconnection, the filament erupts, and an X-point, connecting the postflare loop, appears behind it (see Figure 5(c)). This shows analogy with flare reconnection (see K. Shibata et al. 1995 for details), suggesting this reconnection is the dominant mechanism at this stage. This reconnection process accounts for the more explosive features observed in Figures 2(d), (e), and 3(d)–(f), including the filament eruption, a more violent jet, and a high-energy loop-top source.

The mechanism driving the ascent of the filament, which occurs closely following the interchange reconnection and is visible in Figures 2(a), (b) and 4(d), remains unclear. Our data analysis, combined with previous studies of jets (e.g., K. Shibata et al. 1994; R. L. Moore et al. 2010; E. Pariat et al. 2016; P. F. Wyper et al. 2017, 2018; Y. Chen et al. 2026), led us to propose two possible scenarios to explain this ascent. Both scenarios share the same preconfiguration but differ in their triggers.

The first possibility is that interchange reconnection occurs between the envelope field and the ambient field. In this scenario, interchange reconnection removes the confined field overlying the filament and facilitates its rise. Although this scenario is not depicted in Figure 5, it has been widely seen in the literature (e.g., R. L. Moore et al. 2010; A. C. Sterling et al. 2015; E. Pariat et al. 2016; P. F. Wyper et al. 2017, 2018; Y. Chen et al. 2024, 2026). In such a scenario, the filament rises as an indirect result of the interchange reconnection.

The second possibility, depicted in Figure 5(b), is that the filament is directly involved in interchange reconnection, bypassing its envelope field. In this scenario, the filament reconfigured by interchange reconnection becomes highly curved and tends to uncurl upward in order to relax its magnetic tension (the so-called slingshot effect). During this relaxation, the filament's ascent is attributed to the velocity component of the upward relaxation perpendicular to the solar surface. In such a scenario, the filament rises as a direct result of the interchange reconnection, differing from the first possibility.

These two possibilities can be distinguished by the timing of interchange reconnection relative to the opening process of the filament. Here, we simply assume the appearance time of the Y-point as the occurrence time of interchange reconnection. If the first possibility is at play, the Y-point should be observed prior to the filament opening. This scenario has been clearly demonstrated in previous studies (e.g., F. Fang et al. 2014; P. F. Wyper et al. 2017, 2018; L. Yang et al. 2024; Y. Chen et al. 2026; Y. Duan et al. 2026b). In the current study, however, the Y-point coexists with the filament opening (see Figures 1(d) and (e)), which leads us to consider the second possibility as a more suitable explanation for the observed filament ascent.

Under the framework depicted in Figure 5, it is not necessary for the jet evolution to enter the flare reconnection stage; an alternative pathway exists. Consider the counterbalance between the rising filament and the arresting force that the envelope field exerts on it. The filament erupts, and flare reconnection is triggered if the rising filament overcomes the arresting force of the overlying envelope field. Conversely, if the filament's rise is arrested, leading to its failure to erupt, the jet consequently cannot enter the flare reconnection stage, and only a moderate jet forms.

## 4. Discussions and Conclusions

Using multiwavelength observations from the SDO, we present a detailed analysis of a filament-induced jet. The event proceeds through two distinct reconnection stages: interchange reconnection followed by a more intense flare reconnection, with their morphologies characterized by the transition from a Y-point to an X-point. They also exhibit distinct dynamic properties, including differences in jet speed, maximum height, GOES flux enhancement, and plasma temperature. This study offers the following insights.

*Micro-CME.* CMEs are among the most energetic eruptive activities in the solar atmosphere, driven by flare reconnection (i.e., classical "CSHKP" standard flare model; H. Carmichael 1964; P. A. Sturrock 1966; T. Hirayama 1974; R. A. Kopp & G. W. Pneuman 1976). The jet model proposed by R. L. Moore et al. (2010) extended the picture of CME-related flare reconnection to filament eruptions in blowout jets, known as micro-CMEs. P. F. Wyper et al. (2017) further demonstrated that large-scale CMEs and small-scale filament-induced jets share a universal mechanism. Indirect support came from numerous observational studies. J. Hong et al. (2011) showed that a filament eruption can successively produce a blowout jet and a micro-CME. C. Liu et al. (2011) reveals the consistency between a blowout jet and an X-ray-emitting source, as well as structures analog to large-scale CMEs, suggesting its nature as a micro-CME. A similar picture was also reported by R. Chandra et al. (2017), who identified a high-energy source in the base of a blowout jet.

However, due to observational viewing angles and resolution limitations, key details are often lost due to compact brightening of jet bases, and direct observational evidence of flare reconnection processes in solar jets, i.e., reconnection beneath the erupting filament (Figure 2), remains scarce. Using SDO/AIA observations, we have uncovered direct morphological evidence of flare reconnection during the jet evolution (Figure 2(g)). The consistency with the presence of a high-energy source is also revealed (Figure 2(h)), showing high similarity to the classical flare reconnection picture (for details, see K. Shibata et al. 1995). These findings not only confirm the mini-CME process occurring in the base of a filament-induced jet, but also support the universality of the pathway proposed by P. F. Wyper et al. (2017), highlighting the scale-invariance of solar filament eruptions.

*Coupling between interchange and flare reconnection.* The studies of jets involving difference reconnection reconnection processes have been widely reported in both observations and numerical simulations (C. Liu et al. 2011; V. Archontis & A. W. Hood 2013; E. Pariat et al. 2016; X. Zhu et al. 2017; S. Mandal et al. 2022; Y. Chen et al. 2026), suggesting a potential link between these two reconnection processes. The breakout model provides a natural framework for understanding this linkage, as the positive feedback between reconnection at the breakout current sheet and the subsequent flare reconnection inherently couples the standard and blowout stages (P. Kumar et al. 2018; P. F. Wyper et al. 2018, 2017; Y. Duan et al. 2026b). Slightly different from the breakout model, we propose an alternative scenario in which the filament directly participates in interchange reconnection, as depicted in Figure 5. The relaxation of the reconnected filament drives its upward motion, which in turn creates the conditions necessary for triggering flare reconnection. Although slightly different, this scenario—in which the filament directly participates in interchange reconnection—also involves coupling between interchange and flare reconnection, mediated by the relaxation of the filament's magnetic tension. Such coupling is widely evidenced in previous jet models (V. Archontis & A. W. Hood 2013; F. Moreno-Insertis & K. Galsgaard 2013; P. F. Wyper et al. 2017, 2018).

The above interpretation is derived based on the fact that no magnetic null point is observed in our event. However, it should be noted that, as shown in the animation of Figure 1, a narrow linear feature appears at 06:15 UT. This yields another potential driving mechanism, namely a breakout eruption. A breakout eruption involves the coupling of two reconnection processes: reconnection at the breakout current sheet, which results from the collapse of a magnetic null point into a dissipative current sheet, and flare reconnection beneath the flux rope. The reconnection at the breakout current sheet in its initial stage, before the flux rope enters the breakout current sheet and triggers explosive reconnection (P. Kumar et al. 2018), is usually weak and can therefore manifest as a narrow jet and the lack of an observed null point (P. Kumar et al. 2018, 2021, 2025), consistent with the narrow linear feature observed at 06:15 UT in the animation of Figure 1. This early weak reconnection occurs before the later Y-point/cusp and may therefore represent the less energetic breakout interchange reconnection expected in the breakout picture.

*Distinct interchange and flare reconnection.* The successive occurrence of interchange and flare reconnection in the present event provides an opportunity to compare their distinct dynamics. Figure 4 indicates that the energy release during the flare reconnection stage is significantly stronger than that during the interchange reconnection stage. As discussed by R. Chandra et al. (2017) and references therein, interchange reconnection (referred to as Bald Patch reconnection in their study) alone cannot produce strong GOES flux enhancement, which is commonly attributed to the release of insufficient twist. A comparison with CME studies in X. L. Yan et al. (2018), Y. Chen et al. (2024), and J. Lin & T. G. Forbes (2000), and references therein, reveals the analogy of jets with CMEs, suggesting that blowout jets can be regarded as miniature versions of CMEs (R. L. Moore et al. 2010; C. Liu et al. 2011; A. C. Sterling et al. 2015). Adopting this concept, we reexamine the significant energy difference between two reconnection stages within the framework of CME eruption models.

In the CME framework, the ejection of a flux rope or plasmoid can significantly enhance the reconnection rate (K. Shibata & S. Tanuma 2001). This conclusion applies to any object capable of ejecting magnetic flux (A. Bhattacharjee et al. 2009; D. A. Uzdensky et al. 2010; X. Yan et al. 2022; Z. Tang 2026), including erupting filaments, which are widely evidenced as a ubiquitous source for magnetic flux/twist conversions and transports in the solar atmosphere (Z. Xue et al. 2016; A. C. Sterling et al. 2020, 2024; X. Yan et al. 2020a, 2020b; L. Yang et al. 2026; C. Zhou et al. 2026). Comparing the two reconnection stages in the present event, we propose that the absence of such a contributor to accelerate reconnection leads to a substantially lower reconnection rate in the interchange reconnection stage compared with the following flare reconnection stage, during which filament eruption is present (see Figures 1(d), (e) and 4(d)). We argue that this difference is the key to understanding their distinct energy releases. This argument is observationally supported by C. Liu et al. (2011), whose measurements show that the reconnection rate in the blowout-jet stage is 1 order of magnitude higher than that in the standard-jet stage. The filament eruption thus leads to a higher reconnection rate, which in turn results in stronger energy release during flare reconnection than during interchange reconnection. This is responsible for the intensified observational indicators during the flare reconnection stage, including the X-ray source, plasma temperature, jet height, jet speed, and GOES flux enhancement in the present event. Consequently, the erupting filament is the key resulting in all the differences observed in Figures 1–4, including the dual-stage jet, dual-stage energy release, and dual-stage reconnection.

*Alfvén-speed-comparable jet.* During the flare reconnection stage, higher-speed components are observed compared to the interchange reconnection stage (see Figure 4(a)). Such high-speed components have been observationally reported by J. W. Cirtain et al. (2007) and D. M. Long et al. (2023), with

D. M. Long et al. (2023) finding coexisting low- and high-speed components in blowout jets. Numerical simulations find that the blowout-jet stage exhibits two distinct velocity components: a low-speed component corresponding to the bulk speed of the plasma and a high-speed component corresponding to the Alfvén speed originating from magnetic untwisting (E. Pariat et al. 2016; Y. Chen et al. 2026). In the present event, we identify a high-speed component of approximately 450 km $s^{-1}$ during the flare reconnection stage, which is comparable to the coronal Alfvén speed. This supports the Alfvén nature of the high-speed component in blowout jets. However, unlike the high-speed components reported by D. M. Long et al. (2023) and Y. Chen et al. (2026), which were visible in high-temperature EUV channels, the high-speed component in our event is observed in the SDO/AIA 304 Å channel, which corresponds to transition-region temperatures.

## Acknowledgments

The authors wish to thank the science teams of the Solar Dynamics Observatory (SDO) for their open data policy and the high-quality observations used herein, and the anonymous referee for their insightful suggestions, which have significantly strengthened this work. Data were acquired by GONG instruments operated by NISP/NSO/AURA/NSF with contribution from NOAA. This work is sponsored by the National Key R&D Program of China under No. 2024YFA1612001, the Strategic Priority Research Program of the Chinese Academy of Sciences, grant No. XDB0560000, the National Science Foundation of China (NSFC) under Nos. 12325303, 12473059, 12433010, and 12403065, Yunnan Key Laboratory of Solar Physics and Space Science under No. 202205AG070009, the Yunnan Science Foundation of China under Nos. 202501AW070002 and 202601AS070076, the Yunnan Province Xing Dian Talent Support Program, the Yunling Scholar Project, and the Yunnan Revitalization Talent Support Program under Nos. 202305AS350029, 202305AT350005.

## ORCID iDs

Zehao Tang (汤泽浩) https://orcid.org/0000-0003-0880-9616
Xiaoli Yan (闫晓理) https://orcid.org/0000-0003-2891-6267
Zhike Xue (薛志科) https://orcid.org/0000-0002-6526-5363
Jincheng Wang (王金成) https://orcid.org/0000-0003-4393-9731
Zhe Xu (徐喆) https://orcid.org/0000-0002-9121-9686

## References

Archontis, V., & Hood, A. W. 2013, ApJL, 769, L21
Bale, S. D., Drake, J. F., McManus, M. D., et al. 2023, Natur, 618, 252
Bhattacharjee, A., Huang, Y.-M., Yang, H., & Rogers, B. 2009, PhPl, 16, 112102
Brueckner, G. E., & Bartoe, J.-D. F. 1983, ApJ, 272, 329
Carmichael, H. 1964, NASSP., 50, 451
Chandra, R., Mandrini, C. H., Schmieder, B., et al. 2017, A&A, 598, A41
Chen, B., Kong, X., Yu, S., et al. 2024, ApJ, 971, 85
Chen, Y., Hu, J., Cheng, G., et al. 2024, ApJL, 977, L26
Chen, Y., Peter, H., Przybylski, D., Chitta, L. P., & Mandal, S. 2026, A&A, 706, A141
Chen, Y., Ye, J., Mei, Z., et al. 2022, ApJ, 933, 148
Cheng, G., Ni, L., Tang, Z., et al. 2024, ApJL, 966, L29
Cheng, X., Zhang, J., Saar, S. H., & Ding, M. D. 2012, ApJ, 761, 62
Chitta, L. P., Zhukov, A. N., Berghmans, D., et al. 2023, Sci, 381, 867
Cirtain, J. W., Golub, L., Lundquist, L., et al. 2007, Sci, 318, 1580
Duan, Y., Yan, X., Gao, Y., et al. 2026a, ApJL, 1003, L6
Duan, Y., Yan, X., Hong, J., et al. 2026b, A&A, 706, A1
Fang, F., Fan, Y., & McIntosh, S. W. 2014, ApJL, 789, L19
Forbes, T. G., & Priest, E. R. 1983, SoPh, 84, 169
Guo, F., & Giacalone, J. 2012, ApJ, 753, 28
Hannah, I. G., & Kontar, E. P. 2012, A&A, 539, A146
Harvey, J. W., Bolding, J., Clark, R., et al. 2011, BAAS, 42, 17.45
Hirayama, T. 1974, SoPh, 34, 323
Hong, J., Jiang, Y., Yang, J., Li, H., & Xu, Z. 2017, ApJ, 835, 35
Hong, J., Jiang, Y., Zheng, R., et al. 2011, ApJL, 738, L20
Hong, J., Yang, J., Bi, Y., & Yang, B. 2025, ApJ, 986, 197
Jiang, C., Feng, X., Liu, R., et al. 2021, NatAs, 5, 1126
Joshi, N. C., Nishizuka, N., Filippov, B., Magara, T., & Tlatov, A. G. 2018, MNRAS, 476, 1286
Joshi, R., Schmieder, B., Aulanier, G., Bommier, V., & Chandra, R. 2020, A&A, 642, A169
Karpen, J. T., Antiochos, S. K., & DeVore, C. R. 2012, ApJ, 760, 81
Kopp, R. A., & Pneuman, G. W. 1976, SoPh, 50, 85
Kumar, P., Karpen, J. T., Antiochos, S. K., et al. 2018, ApJ, 854, 155
Kumar, P., Karpen, J. T., Antiochos, S. K., et al. 2021, ApJ, 907, 41
Kumar, P., Karpen, J. T., Wyper, P. F., et al. 2025, ApJ, 994, 180
Kumar, P., Karpen, J. T., Yurchyshyn, V., DeVore, C. R., & Antiochos, S. K. 2024, ApJ, 973, 74
Lemen, J. R., Title, A. M., Akin, D. J., et al. 2012, SoPh, 275, 17
Li, L., Zhang, J., Peter, H., et al. 2016, NatPh, 12, 847
Lin, J., & Forbes, T. G. 2000, JGR, 105, 2375
Lin, R. P., Dennis, B. R., Hurford, G. J., et al. 2002, SoPh, 210, 3
Liu, C., Deng, N., Liu, R., et al. 2011, ApJL, 735, L18
Liu, J., Song, A., Jess, D. B., et al. 2023, ApJS, 266, 17
Liu, M., Ni, L., Cheng, G.-C., Ziegler, U., & Lin, J. 2023, RAA, 23, 035006
Liu, W., Chen, Q., & Petrosian, V. 2013, ApJ, 767, 168
Liu, Y., Shen, J., Yang, X., et al. 2026, ApJ, 1001, 214
Long, D. M., Chitta, L. P., Baker, D., et al. 2023, ApJ, 944, 19
Mandal, S., Chitta, L. P., Peter, H., et al. 2022, A&A, 664, A28
Masuda, S., Kosugi, T., Hara, H., Tsuneta, S., & Ogawara, Y. 1994, Natur, 371, 495
Moore, R. L., Cirtain, J. W., Sterling, A. C., & Falconer, D. A. 2010, ApJ, 720, 757
Moore, R. L., Sterling, A. C., Falconer, D. A., & Robe, D. 2013, ApJ, 769, 134
Moreno-Insertis, F., & Galsgaard, K. 2013, ApJ, 771, 20
Moreno-Insertis, F., Hansteen, V. H., & Nóbrega-Siverio, D. 2025, ApJ, 995, 137
Ni, L., Chen, Y., Peter, H., Tian, H., & Lin, J. 2021, A&A, 646, A88
Nishizuka, N., Shimizu, M., Nakamura, T., et al. 2008, ApJL, 683, L83
Nóbrega-Siverio, D., Joshi, R., Sola-Viladesau, E., Berghmans, D., & Lim, D. 2025, A&A, 702, A188
Ohyama, M., & Shibata, K. 1998, ApJ, 499, 934
Paraschiv, A. R., Bemporad, A., & Sterling, A. C. 2015, A&A, 579, A96
Pariat, E., Dalmasse, K., DeVore, C. R., Antiochos, S. K., & Karpen, J. T. 2015, A&A, 573, A130
Pariat, E., Dalmasse, K., DeVore, C. R., Antiochos, S. K., & Karpen, J. T. 2016, A&A, 596, A36
Priest, E. R., Chitta, L. P., & Syntelis, P. 2018, ApJL, 862, L24
Schou, J., Scherrer, P. H., Bush, R. I., et al. 2012, SoPh, 275, 229
Shibata, K., Masuda, S., Shimojo, M., et al. 1995, ApJL, 451, L83
Shibata, K., Nitta, N., Strong, K. T., et al. 1994, ApJL, 431, L51
Shibata, K., Nakamura, T., Matsumoto, T., et al. 2007, Sci, 318, 1591
Shibata, K., & Tanuma, S. 2001, EP&S, 53, 473
Shimojo, M., & Shibata, K. 2000, ApJ, 542, 1100
Sterling, A. C., Moore, R. L., Falconer, D. A., & Adams, M. 2015, Natur, 523, 437
Sterling, A. C., Moore, R. L., & Panesar, N. K. 2022, ApJ, 927, 127
Sterling, A. C., Moore, R. L., Panesar, N. K., et al. 2020, ApJ, 889, 187
Sterling, A. C., Panesar, N. K., & Moore, R. L. 2024, ApJ, 963, 4
Sturrock, P. A. 1966, Natur, 211, 695
Tang, Z. 2026, ApJ, 1001, 11
Tang, Z., Shen, Y., Liu, D., et al. 2025a, ApJL, 990, L16
Tang, Z., Shen, Y., Zhou, C., Yao, S., & Liu, D. 2025b, ApJL, 985, L11
Tian, H., DeLuca, E. E., Cranmer, S. R., et al. 2014, Sci, 346, 1255711
Tian, H., Yurchyshyn, V., Peter, H., et al. 2018, ApJ, 854, 92

Török, T., Aulanier, G., Schmieder, B., Reeves, K. K., & Golub, L. 2009, ApJ, 704, 485
Uzdensky, D. A., Loureiro, N. F., & Schekochihin, A. A. 2010, PhRvL, 105, 235002
Wang, Y., Ni, L., Cheng, G., et al. 2025, ApJ, 987, 148
White, S. M., Thomas, R. J., & Schwartz, R. A. 2005, SoPh, 227, 231
Wyper, P. F., Antiochos, S. K., & DeVore, C. R. 2017, Natur, 544, 452
Wyper, P. F., DeVore, C. R., & Antiochos, S. K. 2018, ApJ, 852, 98
Xue, Z., Yan, X., Cheng, X., et al. 2016, NatCo, 7, 11837
Yan, X., Li, Q., Chen, G., et al. 2020a, ApJ, 904, 15
Yan, X., Xue, Z., Cheng, X., et al. 2020b, ApJ, 889, 106
Yan, X., Xue, Z., Jiang, C., et al. 2022, NatCo, 13, 640
Yan, X. L., Yang, L. H., Xue, Z. K., et al. 2018, ApJL, 853, L18
Yang, J., Chen, H., Hong, J., Yang, B., & Bi, Y. 2024, ApJ, 964, 7
Yang, J., Jiang, Y., Xu, Z., Bi, Y., & Hong, J. 2015, ApJ, 803, 68
Yang, L., Xue, Z., Wang, J., et al. 2024, ApJ, 976, 135
Yang, L., Yan, X., Wang, J., et al. 2026, ApJ, 1000, 4
Zeng, Z., Chen, B., Ji, H., Goode, P. R., & Cao, W. 2016, ApJL, 819, L3
Zhang, Q. M., & Ji, H. S. 2014, A&A, 567, A11
Zhou, C., Xia, C., Fu, W., Chen, H., & Li, Q. 2026, ApJL, 1006, L39
Zhu, X., Wang, H., Cheng, X., & Huang, C. 2017, ApJL, 844, L20